\documentclass[aps,prx,twocolumn,preprintnumbers,amsmath,amssymb,superscriptaddress,longbibliography,floatfix]{revtex4-2}

\usepackage{xr}

\usepackage[pdftex]{graphicx}
\usepackage{amsfonts}
\usepackage{amssymb}
\usepackage{graphicx}
\usepackage{color}
\usepackage{amsmath}
\usepackage{float}
\usepackage{hyperref}
\usepackage{gensymb}

\usepackage{dcolumn}
\usepackage{bm}
\usepackage{hyperref}

\newcommand{\sdensity}{\widetilde{\chi}(\mathbf{k})}

\begin{document}


\title{Germanium-based nearly hyperuniform nanoarchitectures by ion beam impact}

\author{Jean-Benoit Claude}
\affiliation{Aix Marseille Univ, CNRS, Centrale Marseille, Institut Fresnel, 13013 Marseille, France}
\author{Mohammed Bouabdellaoui}
\affiliation{Aix Marseille Univ, Universit\'e de Toulon, CNRS, IM2NP, Marseille, France}
%
%
\author{Jerome Wenger}
\affiliation{Aix Marseille Univ, CNRS, Centrale Marseille, Institut Fresnel, 13013 Marseille, France}
\author{Monica Bollani}
\affiliation{Istituto di Fotonica e Nanotecnologie-Consiglio Nazionale delle Ricerche, Laboratory for Nanostructure Epitaxy and Spintronics on Silicon, Via Anzani 42, 22100 Como, Italy.}
\author{Marco Salvalaglio} 
\affiliation{Institute  of Scientific Computing,  TU  Dresden,  01062  Dresden,  Germany}
\affiliation{Dresden Center for Computational Materials Science (DCMS), TU  Dresden,  01062  Dresden,  Germany}
%
%
\author{Marco Abbarchi}
\affiliation{Aix Marseille Univ, Universit\'e de Toulon, CNRS, IM2NP, Marseille, France}
\affiliation{
 Solnil, 95 Rue de la R\'epublique, Marseille 13002, France}

\date{\today}

\begin{abstract}
We address the fabrication of nano-architectures by impacting thin layers of amorphous Ge deposited on SiO$_{2}$ with a Ga$^{+}$ ion beam and investigate the structural and optical properties of the resulting patterns. By adjusting beam current and scanning parameters, different classes of nano-architectures can be formed, from elongated and periodic structures to disordered ones with a footprint of a few tens of nm. The latter disordered case features a significant suppression of large length scale fluctuations that are conventionally observed in ordered systems and exhibits a nearly hyperuniform character, as shown by the analysis of the spectral density at small wave vectors. It deviates from conventional random fields as accounted for by the analysis of Minkowski functionals. A proof of concept for potential applications is given by showing peculiar reflection properties of the resulting nano-structured films that exhibit colorization and enhanced light absorption with respect to the flat Ge layer counterpart (up to one order of magnitude at some wavelength). This fabrication method for disordered hyperuniform structures does not depend on the beam size. Being ion beam technology widely adopted in semiconductor foundries over 200 mm wafers, our work provides a viable pathway for obtaining disordered, nearly-hyperuniform materials by self-assembly with a footprint of tens of nanometers for electronic and photonic devices, energy storage and sensing.  
\end{abstract}

\maketitle


\section{\label{sec:Introduction } Introduction }

Hyperuniform (HU) systems are characterized by the suppression of large-wavelength density fluctuations owing to a hidden order which is not apparent on short lengthscales~\cite{Torquato2003,TORQUATO20181}. Systems possessing long-range translational and orientational order, e.g., crystalline structures and periodic fields, represent trivial cases of HU structures. More interestingly, a HU character may be present in disordered systems with partial or full suppression of density fluctuations for small wavevectors (similarly to crystalline solids) and isotropic behaviors with respect to spatial directions (as for liquids). These features are well described for point patterns by the structure factor $S(\mathbf{k})$ with $\mathbf{k}$ the wavevector. Ideal HU characters corresponds to have a scaling $S(\mathbf{k})\sim |\mathbf{k}|^\alpha$ for $|\mathbf{k}|\rightarrow 0$ and $\alpha > 0$. Generalizations have been proposed to account for such a property in systems as, e.g., heterogeneous media as well as scalar and vector fields \cite{Torquato2016,Ma2017,TORQUATO20181}. In experimental settings, quantification of the HU character (how close a system is to the ideal HU case) is typically considered through the metrics $H=S(|\mathbf{k}|\rightarrow 0)/\max(S(|\mathbf{k}|))$, with condition $H\lesssim 10^{-2}$ and $H\lesssim 10^{-4}$ commonly denoting nearly and effective HU character, respectively \cite{Torquato2006,Kim2018}.

\begin{figure*}[ht!]
   \includegraphics[width=1\textwidth]{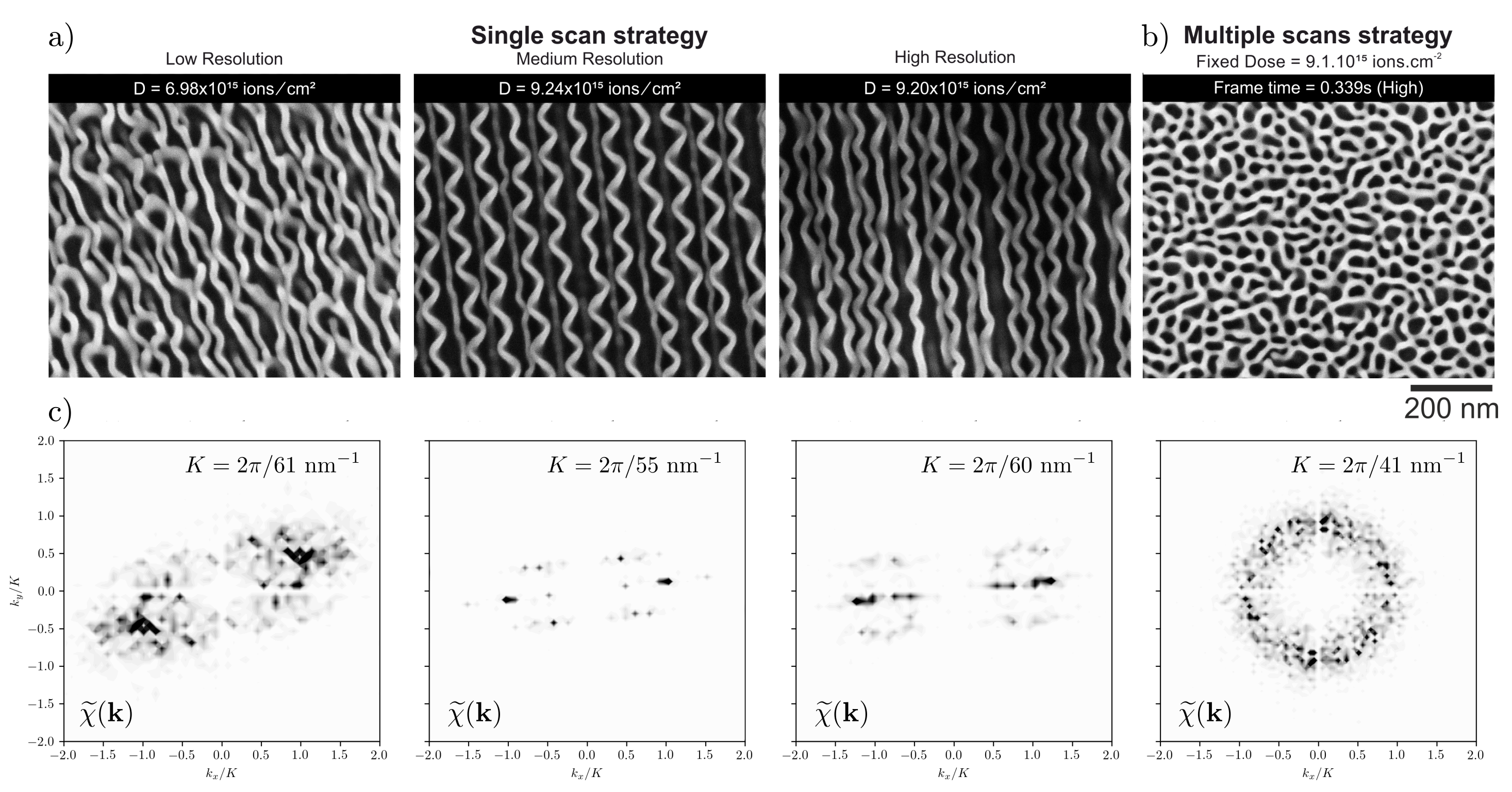} 
   \caption{
    \textit{Influence of FIB single scanning mode}. a) High-resolution scanning electron microscopy (SEM) images of a Ge layer on SOI processed by FIB with a single scan strategy using low, medium, and high resolution (respectively 256 × 221 pixels, 512 × 442 pixels and 1024 × 884 pixels) with ion beam fluency in the 10$^{15}$ ions cm$^{-2}$ range. b) SEM image of a Ge layer on SOI processed by FIB with a multiple scan strategy. In this condition, high and low resolution exhibit the same final morphology with a nearly-HU character. All these patterns above were fabricated on a stack as follows: the structure is composed of a top amorphous germanium layer on a SOI substrate with a top thin silicon layer of 7~nm and a 2000~nm thick SiO$_{2}$ for the BOX. (c) Spectral density $\widetilde{\chi}(\mathbf{k})$ for the samples illustrated in panels (a) and (b). The white-to-black color scale is saturated at $0.75\max(\sdensity)$. $K$ corresponds to the wavevector length at which the $\max(\sdensity)$ is obtained.}
    \label{fig:figure1}
\end{figure*}

Materials and architectures possessing a nearly, effective or ideal HU character have gained a lot of interest in recent years. In addition to fundamental physical aspects~\cite{GAB2002,TORQUATO20181}, their importance
relies in the novel possibilities in light~\cite{FRO2017,ZHO2016,yu2020,FLO2009,MAN2013a,FRO2016,wu2017} and carrier management with topologically-protected electronic states~\cite{MIT2018} as well as vortices in superconductors \cite{rumi2019}. A crucial aspect consists of finding a proper fabrication method that allows the realization of a HU character, e.g., small enough $H$, at wavelengths of interest for the targeted applications. Design protocols for the realization of HU patterns have been developed \cite{FLO2009}. They have usually been combined with top-down approaches. Although powerful, see e.g. \cite{CAS2017,granchi2022near}, such approaches usually show poor scalability. This becomes crucial for patterns meant for large-area applications with features on the nanometer scale, such as metals and nanostructured semiconductor films for photonics applications.

Devising scalable approaches for the fabrication of HU materials featuring nanostructured architectures represents then an important goal. Self-assembly paradigms or hybrid bottom-up/top-down approaches have been exploited for decades to obtain nano- and micro-structures from film-like settings \cite{Medeiros-Ribeiro1998,Ye2011,Naffouti2017}. Self-assembly based on spinodal-like, solid-state dewetting has also recently proved successful for the fabrication of effectively HU, SiGe-based nano-architectures with a footprint of about 100~nm~\cite{salvalaglio2020}. Such an approach can be scaled-up to large surfaces~\cite{benali2020}. Moreover, soft nano-imprint lithography of sol-gel materials (such as silica and titania) was also exploited to fabricate nearly-HU metasurfaces using dewetted micro-and nano-architectures as hard masters \cite{chehadi2021}. 

A remarkable example of self-assembly in crystalline and amorphous Ge semiconductor thin-films consists of the impact of ion beams, widely investigated both from experimental~\cite{bischoff2011,kolibal2011low,rudawski2013,bottger2013} and theoretical point of views~\cite{bradley1988,nord2002,hartmann2009}.
Formation of periodic, ordered~\cite{wei2009,fritzsche2012} and disordered nano-architectures with this approach has been largely addressed so far, and we thus refer the reader to the existing literature~\cite{bradley1988,cuerno1995,castro2005,kolibal2011low}. Central aspects of this approach are its versatility, as the features can be tuned by varying the ion beam parameters (e.g., ion species, energy, and dose ~\cite{bottger2013,hooda2016}, incidence angle~\cite{mollick2014}, sample temperature~\cite{bischoff2011} and stochiometry~\cite{alkhaldi2016}) and processable surface that can reach 200 mm wafers. These structures can also extend over several tens of micrometers in depth, below the surface, resulting in actual three-dimensional, porous nanoarchitectures with an average cavity radius as small as a few tens of nm~\cite{romano2012,alkhaldi2016}.

Here we report on Ge-based nano-structures obtained by the impact of a high-energy, Ga$^{+}$ ion beam on Ge layers resulting in ordered and disordered nano-architectures and characterize their morphological features. In Sect.~\ref{sec:experiments} we show the formation of strongly anisotropic and wavy structures as well as isotropic disordered ones, obtained by adjusting ion current, supplied dose, and scanning parameters. We focus then on the isotropic disordered pattern, being the main target of this study, and we discuss its features in Sect.~\ref{sec:experiments}. In particular, the nearly-HU character of the disordered structures is assessed through the analysis of their spectral density that accounts for the formation of Ge-based nearly-HU nano-architectures having a record footprint of about 40~nm. The analysis via Minkowski functionals further assesses the deviations from a (Gaussian) random field and supports the presence of correlations in the patterns. 
A proof of concept for potential applications is then given in Sect.~\ref{sec:application}, showing peculiar reflection properties of the resulting nanostructured film. For the sake of readibility, we report in the main text the main evidence and results, while including additional materials in the Supporting Information.
Conclusions are summarized in Sect.~\ref{sec:conclusions}

\section{Experiments}
\label{sec:experiments}

Experiments were performed on a 50-nm thick amorphous germanium layer deposited on a silicon-on-insulator (SOI) substrate composed of a top silicon layer of 7~nm on a 2000~nm thick SiO$_{2}$ layer. After chemical cleaning (5 seconds in 10$\%$ HF solution in H$_{2}$O in N$_{2}$ atmosphere), amorphous Ge is deposited by molecular beam epitaxy (Riber R32, in ultra-high vacuum of $\sim$10$^{-10}$ torr, at room temperature and growth rate of 10~\AA/min). 

Although very similar structures can be obtained on crystalline bulk Ge (Supplementary Information Figure 5), we chose to implement them in ultra-thin Ge layers on commercial ultra-thin SOI. This substrate is more suitable for electronic and photonic device integration and is thus more appealing than the bulk counterpart. For the sake of thoroughness, we mention that several other experimental conditions were explored, such as a-Ge 80~nm thick deposited atop SOI (Supplementary Information Figure 5) or custom-made germanium on insulator wafers with 200~nm a-Ge atop 300~nm of SiO$_2$ (not shown). They always provided similar results to those shown here, accounting for the possibility of fabricating these structures in different conditions. 

For this study, we used a dual-beam FEI Strata DB235 with Ga$^{+}$ (atomic mass 31 amu) liquid-metal ion source focused ion beam (FIB) and a SFEG source for the scanning electron microscope imaging (SEM). We used ion currents within 15 and 1000~pA at 30~keV. The samples are kept orthogonal to the Ga ion beam. The FIB beam moves horizontally, from the top left part of the window and line-by-line, raster-scans the sample surface, down to the bottom right point. 

For single scans (Figure~\ref{fig:figure1} a), the ion beam resolution is set to low (256 $\times$ 221 pixels), medium (512 $\times$ 442 pixels), and high (1024 $\times$ 884). At 50000$\times$ magnification (used for the milling process), the field of view has a side of 6080~nm. Thus, each pixel is respectively separated by 24, 12, and 6~nm for low, medium, and high resolution. The ion current was fixed at 30~pA, corresponding to a beam diameter of about 15~nm. Thus, only low-resolution modes do not result in a beam overlapping from one point to the other. As shown in Figure\ref{fig:figure1} a), different resolution impacts the final morphology with different features. 

In analogy with what is found under strong laser illumination\cite{HER1998} and ion beam impact in metals\cite{repetto2012,repetto2015}, several factors are expected to contribute to the formation of ordered and disordered patterns, namely ion-induced melting of the a-Ge layer, curvature-dependent sputtering and/or swelling \cite{yanagisawa2007}. For Ge layers, a relevant role in the pattern formation was attributed to vacancy clustering \cite{rudawski2013}. Temperature and mass redistribution at surfaces are also expected to play a role as supported by observations in similar systems\cite{wei2009} and by the theory of pattern formation by ion bombardment \cite{bradley1988,bradley2010,cuerno1995,castro2005}. A comprehensive investigation and tuning of all these contributions are beyond the scope of this work, which instead focuses on selected conditions leading to interesting and potentially relevant properties, as discussed in the following.

\begin{figure}[t]
   \includegraphics[width=0.9\columnwidth]{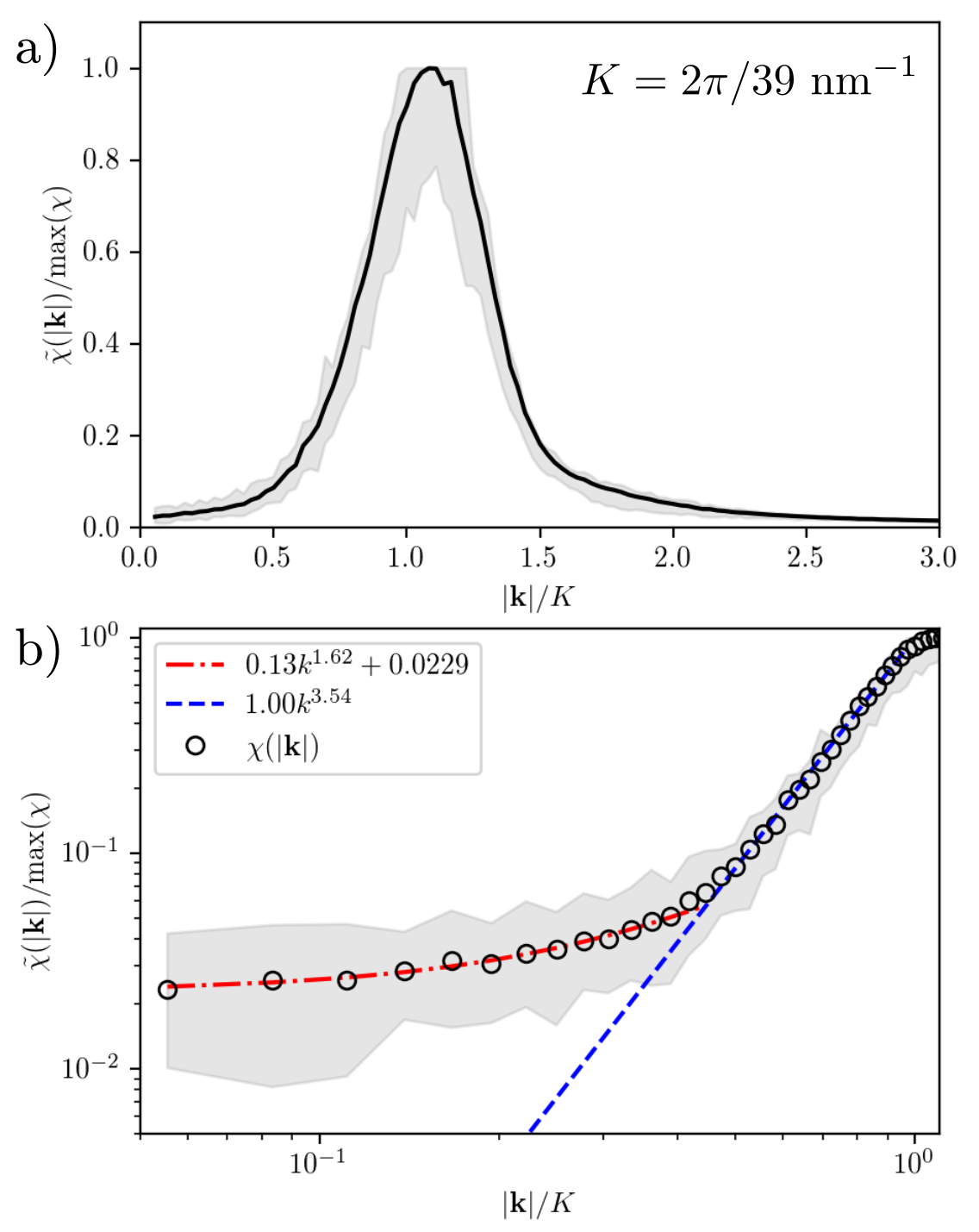} 
   \caption{
   \textit{Spectral density and assessment of the nearly-HU character}. a) Radial distribution of the normalized spectral density for a porous a-Ge sample averaged over ~50 images of different location in the same sample (see the Supplementary Information Figure 8). The shaded grey areas displays the variability of the spectral density measured over all the 50 repetitions of the pattern. b) Log-log plot for the curve in panel (a) focusing on the region $|\mathbf{k}|/K\leq1$ and showing the decay for $|\mathbf{k}|\rightarrow 0$.}
    \label{fig:figure3}
\end{figure}

\begin{figure}[t]
   \includegraphics[width=1\columnwidth]{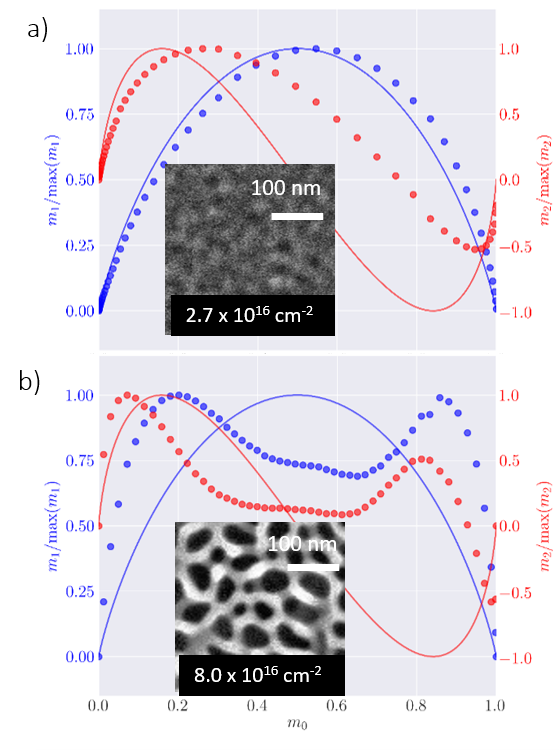} 
   \caption{
   \textit{Disorder and correlations of spontaneous patterns}. a)  Minkowski functionals analysis $m_{1,2}(m_{0})$ for a 30~nm thick Ge layer exposed to multiple scans with a  2.7$\times$10$^{15}$ ions cm$^{-2}$ dose.  b) Same as a) for 8.02$\times$10$^{15}$ ions cm$^{-2}$ dose. Continuous lines represent the trend of $m_{1,2}(m_{0})$ for a Gaussian random field. The insets show SEM micrographs with details of the sample's surface. }
    \label{fig:figure2}
\end{figure}

\section{Pattern analysis}

Global features of the patterns as in Figure~\ref{fig:figure1} can be analyzed by looking at their spectral density $\widetilde{\chi}(\mathbf{k})$. It provides the generalization for the information conveyed by the structure factor $S(\mathbf{k})$ for point patterns \cite{TORQUATO20181}. The SEM images collected for the structures (c.f. Figure~\ref{fig:figure1}) well resolve the region occupied by the solid phase with respect to the surrounding vacuum. Therefore we consider the spectral density $\widetilde{\chi}(\mathbf{k})$ for a two-phase (solid-vacuum) system \cite{Torquato2016}. It is based on an indicator function:
\begin{equation}
\mathcal{I}^{(s)}(\mathbf{r})=
    \begin{cases}
    1, & \mathbf{r}\in \mathcal{V}_s, \\
    0, & \mathbf{r}\notin \mathcal{V}_s,
    \end{cases}
\end{equation}
with $\mathbf{r}\in \Omega$ the spatial coordinates and $\mathcal{V}_s$ the region with grey-scale values above a given threshold. $\widetilde{\chi}(\mathbf{k})$ then corresponds to the Fourier transform of the autocovariance function: \cite{torquato1999exact,Torquato2016}
\begin{equation}
    \chi(\mathbf{r})=\mathcal{S}_2^{(s)}(\mathbf{r})-(\phi^{(s)})^2,
\end{equation}
with
\begin{equation}\label{eq:Ix}
    \mathcal{S}_2^{(s)}(\mathbf{r})=\langle \mathcal{I}^{(s)}(\mathbf{r}')\mathcal{I}^{(s)}(\mathbf{r}'+\mathbf{r}) \rangle,
\end{equation}
the two-points autocorrelation function and $\phi^{(s)}=\langle \mathcal{I}^{(s)}(\mathbf{r}') \rangle$ the volume fraction of the solid phase. The spectral density can be computed directly as: \cite{torquato1999exact}
\begin{equation}
    \widetilde{\chi}(\mathbf{k})=\frac{1}{|\Omega|}|\mathcal{F}[\mathcal{I}^{(\rm{s})}(\mathbf{r})-\phi_i]|^2
\end{equation}
with $\mathcal{F}$ the Fourier transform and $|\Omega|$ the area of the images. The decay of $\widetilde{\chi}(\mathbf{k})$ for $\mathbf{k}\rightarrow 0$ conveys similar information to $S(|\mathbf{k}|)$ \cite{Torquato2016,TORQUATO20181}\footnote{Interestingly, results similar to the one reported in the following can be obtained by sampling the solid phase with randomly distributed points and computed $S(\mathbf{k})$.}. 

The spectral density for the patterns obtained in Figure~\ref{fig:figure1} a) is shown in the first three panels of Figure~\ref{fig:figure1} c). The structures exhibit anisotropic arrangements with marked periodicities. The characteristic frequency $K$ (where the peak of $\widetilde{\chi}(\mathbf{k})$ is observed) is reported in the figures and points to characteristic length scales $2\pi/K$. For low resolution, the structures present a higher degree of disorder which is then reduced for larger resolution. The patterns are also found to exhibit slightly different predominant orientations $>$83$^\circ$ tilt with respect to the scan direction, medium resolution 75$^\circ$-83$^\circ$, and low resolution 61$^\circ$-70$^\circ$. As expected for (partially-) ordered structures, the corresponding spectral density show an exclusion zone at small wavevectors.  
This analysis shows that, depending on how the ions are supplied, different anisotropic structures can be obtained with tunable disorder. 


A behavior similar to the patterns illustrated above in terms of suppression of correlations at large wavelength (with an exclusion zone at small wavevectors), has been obtained through a multiple scans strategy, using an overall dose of $9.1 \times$10$^{15}$ ions cm$^{-2}$(Figure~\ref{fig:figure1} b). However, this time the final structure features a spectral density $\widetilde{\chi}(\mathbf{k})$ with isotropic angular distribution (last panel in Figure~\ref{fig:figure1} c) that corresponds to a disordered and uniform distribution of objects. Exploring the three different resolution configurations always provides the same morphology featuring connected and disordered structures, even when using an overall dose comparable to that of a single scan (see Supplementary Information Figure 7). 

Beyond raster scans with a small spot (as shown in Figure\ref{fig:figure1} b)), disordered, porous a-Ge can be obtained also by ion impact on a single large spot (see the Supplementary Information Figure 1-4). This demonstrates that the fabrication process can be scaled to much larger surfaces as it does not require a high lateral resolution. Nonetheless, the possibility of creating these patterns over small and well-defined areas may allow to integrate them within other devices.

\begin{figure}[h!]
   \includegraphics[width=0.9\columnwidth]{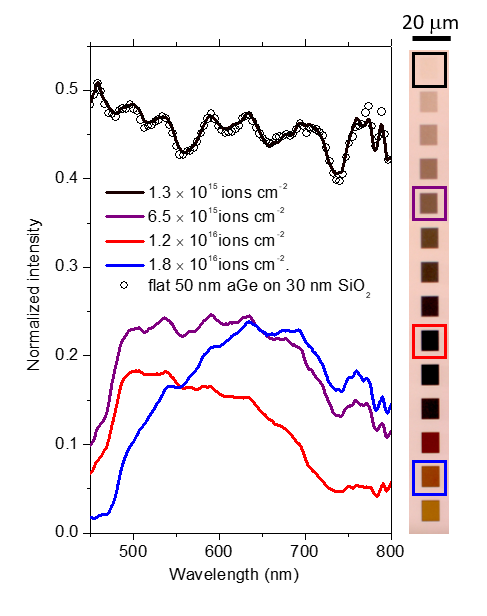} 
   \caption{
   \textit{Light reflection from porous a-Ge}. a) Reflection from 50~nm a-Ge on 30~nm thick SiO$_{2}$ normalized by the white tungsten lamp used for illumination. Open symbols correspond to flat a-Ge, whereas continuous lines are impacted areas with a multiple scan approach. Spectra are collected in bright field illumination with a 100$\times$ magnification objective lens, numerical aperture 0.9 (collection angle about $\pm$64 degrees, lateral, resolution $\sim$6~$\mu$m. Right inset: from top to bottom, bright-field images of the impacted areas with doses ranging from 0 to 2.3$\times$10$^{16}$ ions cm$^{-2}$ in steps of 5$\times$10$^{15}$ ions cm$^{-2}$. The highlighted areas are selected for spectroscopy. }
    \label{fig:figure4}
\end{figure}

We now further asses the disorder found in the case of uniform structures. We provide a quantitative assessment of the HU character of the uniform and disordered pattern in Figure~\ref{fig:figure1} b) and address their scalability by analyzing 50 separate images in different points in the sample (see a collage in the Supplementary Information Figure 8). In Figure~\ref{fig:figure3}, we report the normalized radial distribution of $\sdensity$ averaged over all the analyzed images. Grey areas show the range of variability of the results across the different images. Figure~\ref{fig:figure3}a shows a marked peak at about 0.16~nm$^{-1}$ corresponding to a characteristic length-scale of $\sim 39$~nm ($\pm 4$~nm). We observe a decay of $\sdensity$ toward zero, sampled up to the limits of the analyzed window, well approximated by $|\mathbf{k}|^{\alpha}$ with $\alpha\approx 3.5$ close to $|\mathbf{k}|/K=1$ (see log-log plot in Figure~\ref{fig:figure3}b). 

An extrapolation of $\widetilde{\chi}(0)$ through fitting allows to estimate a H-metric value of $H\approx2\cdot 10^{-2}$ for the averaged $\sdensity$, and within a range $H\in [8 \cdot 10^{-3}-3 \cdot 10^{-2}]$ among the analysed samples. This value is compatible with a nearly-HU character \cite{Torquato2006,Kim2018} similarly to other systems where correlation and HU character emerge spontaneously \cite{wilken2020,xie2013hyperuniformity}.

The features of the disordered structures, beyond simple random fields, can be further accounted for by Minkowski functional analysis describing their topological features~\cite{MIN1989,SCH2014,salvalaglio2020}. We plot Minkowski functionals $m_i(\bar{\rho})=(1/|\Omega|)M_i(\mathcal{B}_{\bar{\rho}})$ %
with 
$\bar{\rho}$ a given threshold defining a binary image of thresholded gray-scale images~\cite{Mantz_2008,salvalaglio2020} (Figure~\ref{fig:figure2}).
%
%
$m_0(\bar{\rho})$ is the fraction of $|\Omega|$ occupied by the non-zero region in $\mathcal{B}_{\bar{\rho}}$, as $M_0(\mathcal{B}_{\bar{\rho}})$ represents the area of  $\mathcal{B}_{\bar{\rho}}=1$. $m_1(\rho)$ is the average of the boundary length $U$ between the areas where $\mathcal{B}_{\bar{\rho}}=1$ and $\mathcal{B}_{\bar{\rho}}=0$, as $U=2 \pi M_1(\mathcal{B}_{\bar{\rho}})$. $m_2(\rho)$ is the averaged Euler characteristic $\chi$, as $\chi=\pi M_2(\mathcal{B}_{\bar{\rho}})$. With this method, the results are independent of image saturation and contrast~\cite{Mantz_2008,salvalaglio2020}. 

$m_{1,2}(m_{0})$ assess the deviation from a random field (solid lines in Figure~\ref{fig:figure3} a) and b)) and eventual non-linearity\cite{bradley1988,alkemade2006} of the underlying pattern formation dynamics~\cite{Mantz_2008}. We observe significant deviations from a Gaussian random field of $m_{1}(m_{0})$ and $m_{2}(m_{0})$ for porous Ge irradiated with 8.0$\times$10$^{15}$ ions cm$^{-2}$ (Figure~\ref{fig:figure3} b)), which increase with increasing the dose. 
Other FIB settings (e.g., lower beam energy), as in T. Bottger et al. \cite{bottger2013} result in disordered structures very close to a Gaussian random field (not shown).

\section{Optical properties}
\label{sec:application}

Finally, we address the changes in light reflection from the impacted areas featuring a nearly-HU character. A spectroscopic characterization of light reflection performed in bright-field illumination reveals strong changes in the affected areas' optical response with respect to the flat counterpart (Figure~\ref{fig:figure4}). The 50~nm thick a-Ge layer deposited atop 30~nm thick SiO$_{2}$ shows a drastic reduction in the reflected light intensity when impacted with a dose larger than $\sim$ 5$\times$10$^{15}$ cm$^{-2}$ in single scan mode. A red colorization appears for larger doses (from about 1.3$\times$10$^{16}$  cm$^{-2}$) owing to enhanced absorption at short wavelength, cutting the blue light reflection (up to about 25 times larger than the flat counterpart). 
Qualitatively similar results have been obtained with 200~nm thick a-Ge, and with a-Ge deposited on thin and thick silicon on insulator (not shown).

Provided the presence of several layers with different optical constants, the overall response to illumination is determined by etalon effects and by the disordered network of filaments~\cite{galinski2017,galinski2017b}. 

\section{Conclusions}
\label{sec:conclusions}
In conclusion, we showed that impacting a thin layer of Ge with an ion beam results in ordered and disordered structures with a nearly-HU character. Their typical size is the smallest reported so far for analogous HU systems~\cite{salvalaglio2020,CAS2017,tavakoli2020}. Provided the possibility to change stoichiometry by Si alloying~\cite{alkhaldi2016}, tune the size of the pores by changing ion dose~\cite{hooda2016,rudawski2013} or Ge film thickness~\cite{romano2012} and also obtain truly 3D structures~\cite{romano2012,rudawski2013}, this fabrication approach has considerable potential for overcoming the limitations of conventional methods for nanoarchitectures with HU character~\cite{soukoulis2011}.


\section*{Acknowledgments}
This research was funded by the EU H2020 FET-OPEN project
NARCISO (No. 828890). J. W. and J.-B. C. acknowledge
the European Research Council (ERC, Grant Agreement
No. 723241). We acknowledge the Nanotecmat platform of the IM2NP institute of Marseilles.

\providecommand{\noopsort}[1]{}\providecommand{\singleletter}[1]{#1}%

\end{document}